\PassOptionsToPackage{sectionbib}{natbib}
\documentclass[%
 reprint,
 amsmath,amssymb,
 aps,
]{revtex4-2}
\usepackage{comment}
\usepackage{graphicx}
\usepackage{siunitx}
\usepackage{blindtext}
\usepackage{newfloat}
\usepackage[resetlabels]{multibib}
\usepackage[hidelinks]{hyperref}
\usepackage{chapterbib}
\usepackage[capitalise]{cleveref}
\usepackage[none]{hyphenat}
\usepackage{xurl}
\setcitestyle{super,open={},close={}}

\newcommand{\md}{\mathrm{d}}
\renewcommand{\figurename}{Fig.}
\DeclareFloatingEnvironment[name={Extended Data Table},within=none]{exttable}
\DeclareFloatingEnvironment[name={Extended Data Fig.},within=none]{extfigure}

\begin{document}


\title{Enhanced singular jet formation in oil-coated bubble bursting}
\author{Zhengyu Yang$^{1\ast}$}
\author{Bingqiang Ji$^{1\ast}$}
\author{Jesse T. Ault$^{2}$}
\author{Jie Feng$^{1,3\dagger}$}%
\affiliation{%
 $^1$Department of Mechanical Science and Engineering, University of Illinois at Urbana–Champaign, Urbana, IL, 61801, USA.}
 \affiliation{%
 $^2$School of Engineering, Brown University, Providence, RI, 02912, USA.}
\affiliation{%
$^3$Materials Research Laboratory, University of Illinois at Urbana–Champaign, Urbana, IL, 61801, USA.}
\affiliation{%
$^\ast$These authors contributed equally to this work. \\
$^\dagger$Email: jiefeng@illinois.edu.
}%
\begin{abstract}
\noindent\normalsize \textbf {Bubbles are ubiquitous in many natural and engineering processes, and bubble bursting aerosols are of particular interest because of their critical role in mass and momentum transfer across interfaces. All prior studies claim that bursting of a millimeter-sized bare bubble at an aqueous surface produces jet drops with a typical size of $\boldsymbol{O}$(100 $\si{\micro\relax}$m), much larger than film drops of $\boldsymbol{O}$(1 $\si{\micro\relax}$m) from the disintegration of a bubble cap. Here, we document the hitherto unknown phenomenon that jet drops can be as small as a few microns when the bursting bubble is coated by a thin oil layer. We provide evidence that the faster and smaller jet drops result from the singular dynamics of the oil-coated cavity collapse. The unique air-oil-water compound interface offers a distinct damping mechanism to smooth out the precursor capillary waves during cavity collapse, leading to a more efficient focusing of the dominant wave and thus allowing singular jets over a much wider parameter space beyond that of a bare bubble. We develop a theoretical explanation for the parameter limits of the singular jet regime by considering the interplay among inertia, surface tension, and viscous effects. As such contaminated bubbles are widely observed, the previously unrecognized fast and small contaminant-laden jet drops may enhance bubble-driven flux across the interface, contributing to the aerosolization and airborne transmission of bulk substances.}
\end{abstract}
\maketitle

\noindent Bubbles present in liquids are commonplace in a wide spectrum of natural and industrial processes \citep{gonnermann2007fluid,bird2010daughter,feng2014,veron2015ocean,dollet2019bubble,oratis2020new,liger2021recent,deike2022mass}. In the cases where bubbles rise to the liquid surface, they burst and generate numerous droplets including jet drops and film drops\citep{anguelova2021big,deike2022mass}. The jet drops are formed by the fragmentation of an upward jet induced by the cavity collapse, while the film drops are produced from the disintegration of the bubble cap.  These drops play a significant role in many transport processes across the air-liquid interface\citep{veron2015ocean,ji2021compound,spiel1995, liger2021recent,ma2020characteristics}. For example, drops from bursting bubbles are considered as a main source of sea spray aerosols\citep{veron2015ocean}, impacting air pollution\citep{murphy2016depth,trainic2020airborne}, global climate\citep{wang2017role, veron2015ocean, wilson2015}, and even infectious disease transmission\citep{ji2022water,bourouiba2020,mcrae2021aerosol,bourouiba2021fluid}. While most prior studies focus on clean bubbles, in practice bubbles with a compound interface are more ubiquitous. Such bubbles could be formed as gas bubbles in natural water bodies scavenge surface-active organic materials while they rise\citep{blanchard1970}. Other examples include gas bubbles released from natural seeps\citep{johansen2017time}, froth flotation\citep{behrens2020oil}, and material processing using coated bubbles\citep{visser2019architected,behrens2020oil}. Within this context, it remains unclear how a compound interface, \textit{e.g.} one that is formed by an insoluble coating at the contaminated bubble surface, mediates the bubble bursting dynamics and the related mass and momentum transport.


\begin{figure*}
    \centering
    \includegraphics[width=0.9\linewidth]{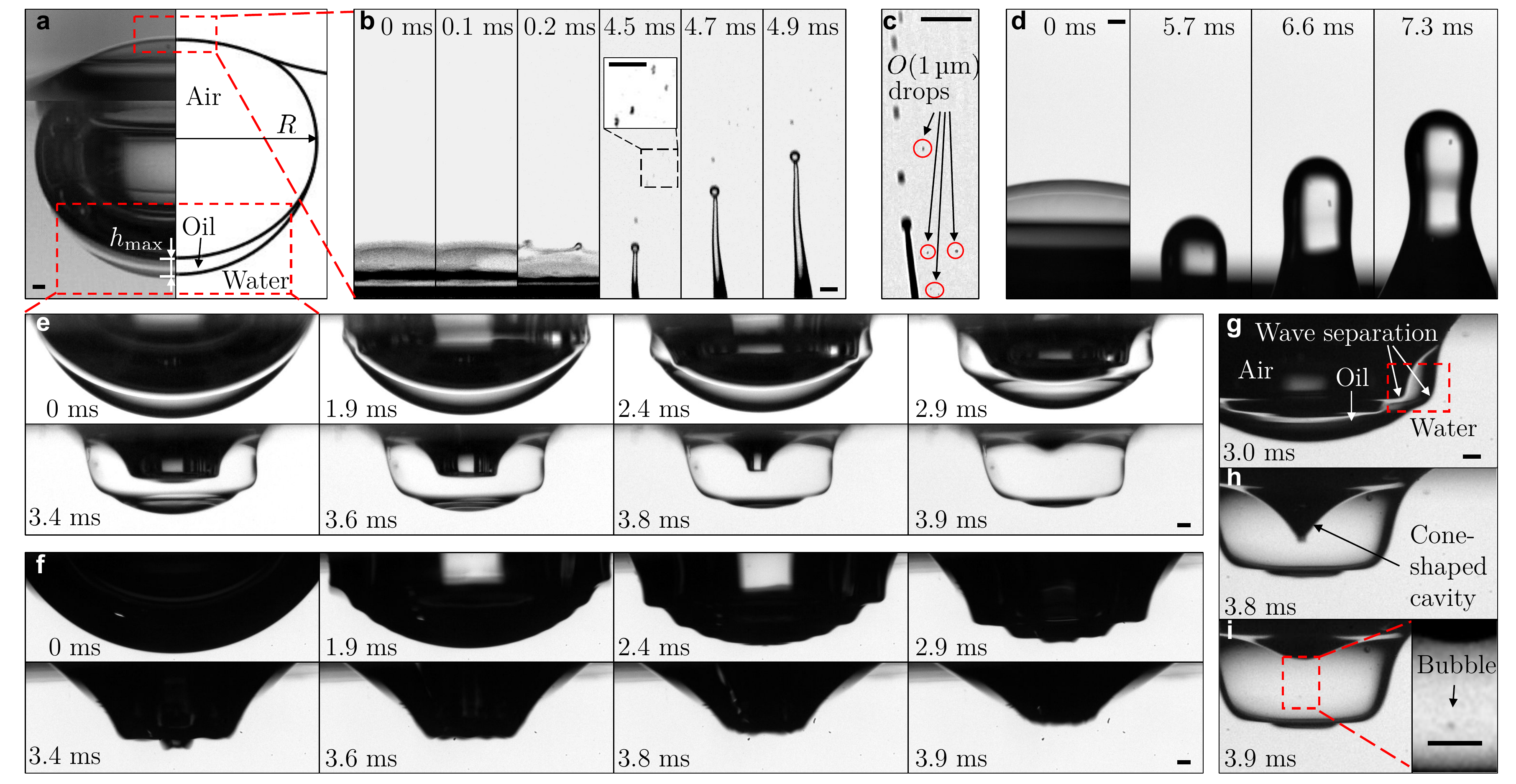}
    \caption{\textbf{Oil-coated bubble bursting at an aqueous surface.} \textbf{a}, Experimental image (left) and schematic (right) of an oil-coated bubble resting on a free aqueous surface. \textbf{b}, High-speed side-view snapshots of film rupturing and jet formation during oil-coated bubble bursting, with oil fraction $\psi_o=10\%$ and oil viscosity $\mu_o=$ \SI{4.6}{\milli\pascal\second}. The small jet drops are highlighted in the inset. $t=0$ represents the moment when a hole nucleates in the bubble cap. \textbf{c}, High-resolution image of jetting during an oil-coated bubble bursting at $t=3.98$ ms, with $\psi_o=3.8\%$ and $\mu_o=$ \SI{4.6}{\milli\pascal\second}. Drops as small as $O(\SI{1}{\micro\meter})$ are highlighted by the red circles. \textbf{d}, Jetting of a bare bubble bursting at a clean water surface. \textbf{e}, Side-view snapshots of an oil-coated cavity collapsing simultaneously to \textbf{b} where $\psi_o=10\%$ and $\mu_o=$ \SI{4.6}{\milli\pascal\second}. \textbf{f}, Cavity collapsing during bare bubble bursting. \textbf{g-i}, Side-view zoom-in snapshots of capillary wave separation (\textbf{g}), cone-shaped cavity (\textbf{h}) and bubble entrapment upon singular jet formation (\textbf{i}). Here, $\psi_o=4.2\%$ and $\mu_o=$ \SI{1.8}{\milli\pascal\second}. The bubble radius is $R=2.1\pm0.3$ mm. All scale bars represent \SI{200}{\micro\meter}.}
    \label{Exp}
\end{figure*}
All prior work claims that a millimeter-sized bare bubble bursting in water produces jet drops of $O(\SI{100}{\micro\meter})$ (with a typical ejection velocity of $O(1\ \text{m/s})$)  \citep{blanco2021jets, lhuissier2012bursting} which are unlikely to contribute to bubble-driven aerosols because of the short floating duration \citep{veron2015ocean}. However, we show that the bursting of a millimeter-sized bubble contaminated by an oil coating (Fig.~\ref{Exp}a-c and Extended Data Fig. \ref{ExpSche}) on clean water can generate micron-sized jet drops with an ejection velocity as large as $O$(10 m/s). We are not aware of any previous documentation of this phenomenon. Figure~\ref{Exp}c shows that such an extremely thin and fast jet emerges above the water surface after bubble cap rupture, and then breaks up into multiple jet drops.
We further confirm that the jet drops consist of oil only when the oil volume fraction (defined as the ratio between the oil and bubble volumes, see Methods) $\psi_o\ge 0.5\%$, by checking the jet drop composition using a test strip for water detection (see Methods and Extended Data Fig.~\ref{fig:droptest}).
In contrast, the same fast and thin jet is not observed for a bare bubble bursting in a pure water or oil phase (Fig.~\ref{Exp}d and Extended Data Fig.~\ref{fig:nonsingular}). Smaller micron-sized oily jet drops are noteworthy because the slower settling velocities allow them to persist longer and travel further, as the film drops with similar sizes\citep{veron2015ocean}. Thus, they unavoidably affect the chemical compositions of the sea spray aerosols\citep{wang2017role} and the airborne transmission of bulk substances such as contaminants and viruses \citep{bourouiba2020,bourouiba2021fluid}. Our findings suggest the role of the jet drops in bubble-bursting aerosols should be carefully revisited for a compound bubble.  

The fast, thin jets observed here are often referred to as singular jets\citep{zeff2000singularity,bartolo2006singular,ganan2017revision}, which are found to result from finite-time self-similar dynamics of the cavity collapse\citep{zeff2000singularity,lai2018bubble}. Side-view high-speed observations show how the oil-coated cavity collapse leads to the formation of a singular jet (Fig.~\ref{Exp}e) distinct from bare bubble bursting (Fig.~\ref{Exp}f). The compound bubble is fully engulfed by silicone oil initially. After the bubble cap ruptures, a train of capillary waves is excited and travels downwards along the air-oil-water interface. As the oil is swept towards the cavity nadir, the oil layer becomes thicker, and the capillary waves separate onto both the air-oil and oil-water interfaces (Fig.~\ref{Exp}g). The capillary waves at the air-oil interface propagate faster than those at the oil-water interface, and finally merge at the cavity nadir, forming a cone-shaped cavity in the oil domain that generates an upward oily jet (Fig.~\ref{Exp}h). Finally, a tiny bubble is trapped in the oil domain (Fig.~\ref{Exp}i), a feature consistent with prior experimental and numerical observations for singular jets produced from bare bubble bursting\citep{duchemin2002jet,gordillo2019capillary}.
\begin{figure*}
    \centering
    \includegraphics[width=0.85\linewidth]{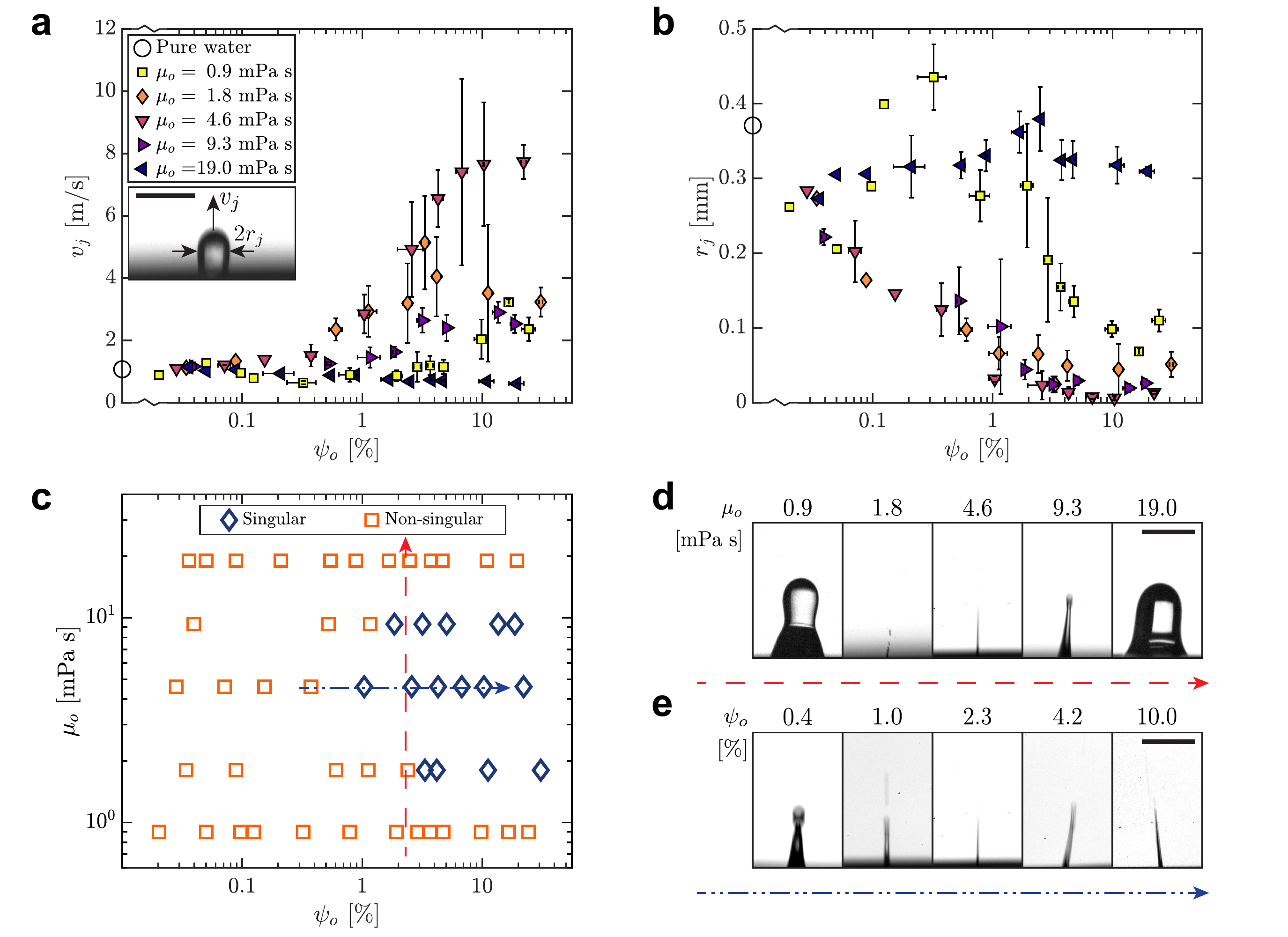}
    \caption{\textbf{Characterization of jets produced from oil-coated bubble bursting.} \textbf{a}, Jet velocity $v_j$ as a function of oil fraction $\psi_o$ at different oil viscosities $\mu_o$. The inset shows the moment when $v_j$ and the jet radius $r_j$ are measured. The scale bar represents 1 mm. \textbf{b}, $r_j$ as a function of $\psi_o$ at different $\mu_o$. The hollow markers at the left vertical axis of \textbf{a-b} represent the case of a bare bubble of the same size ($\psi_o=0\%$) bursting in pure water. Error bars are calculated as the standard deviations of at least three measurements. \textbf{c}, Regime map of jet singularity with $\psi_o$ and $\mu_o$. The red dashed and blue dot-dashed lines correspond to the experimental cases in \textbf{d} and \textbf{e}, respectively. \textbf{d-e}, Experimentally observed jet morphology for different $\mu_o$ with $\psi_o=2.3\pm0.6\%$ (\textbf{d}), and  different $\psi_o$ with $\mu_o=4.6$ mPa s (\textbf{e}). The scale bars represent 1 mm. }
    \label{Regm}
\end{figure*}

For bare bubble bursting, the dimensionless numbers $Oh=\mu/\sqrt{\rho\gamma R}$ (representing the ratio of viscous to inertial and capillary effects) and $Bo = \rho gR^2/\gamma$ (representing the ratio of gravitational to capillary effects, negligible in current experiments) determine the jet dynamics, where $\rho$ is the liquid density, $\gamma$ is the surface tension and $R$ is the bubble radius. No jet singularity is predicted for millimeter-sized bare bubbles bursting in either pure water or silicone oil\citep{brasz2018minimum,deike2018dynamics,gordillo2019capillary,blanco2020sea}, as verified by experiments (Extended Data Table 1 and Extended Data Fig.~\ref{fig:nonsingular}). Nevertheless, in compound bubble bursting, singular jet formation is experimentally observed for a wide range of oil viscosities $\mu_o$ and oil volume fractions $\psi_o$. These singular jettings are characterized by narrow and fast jets with high peak jet velocities sensitive to the initial bubble rupture process\citep{michon2017jet,yang2020multitude,thoroddsen2018singular}. In addition, we note that we experimentally and theoretically obtain that $h_{\max}/R\sim\psi_o^{2/3}$, where $h_{\max}$ is the maximum oil layer thickness at the bottom pole of the bubble. (Fig.~\ref{Exp}a and SI section S3).  

Inspired by previous studies\citep{brasz2018minimum,duchemin2002jet,deike2018dynamics}, we define singular jets as those with a dimensionless tip radius $r_j/R\le0.025$, which corresponds to a dimensionless tip velocity $v_j/v_{ce} \gtrsim14$ (Fig.~\ref{Regm}a-b and Extended Data Fig.~\ref{fig:dimlessext}). Here, $v_{ce}=\sqrt{\gamma_e/(\rho_wR)}$ is the capillary velocity, where the effective surface tension for an air-oil-water interface $\gamma_e=\gamma_{oa}+\gamma_{ow}$ is the sum of the oil-air and oil-water surface tensions. The jet tip velocity $v_j$ and radius $r_j$ are measured when the jet crosses the undisturbed air-water interface (inset of Fig.~\ref{Regm}a). For oil-coated bubbles with $\mu_o = 1.8-9.3$ mPa s, singular jets occur when $\psi_o > 2\%$, as $v_j$ increases rapidly with a sharply decreasing $r_j$. Meanwhile, for $\mu_o = 0.9$ or 19 mPa s, no jet singularity is observed for all $\psi_o$ (Fig. S1). To the best of our knowledge, the evolution of the jet morphology with $\mu_o$ and $\psi_o$ is shown in Fig.~\ref{Regm}c-e for the first time. 



\begin{figure*}
    \centering
    \includegraphics[width=0.85\linewidth]{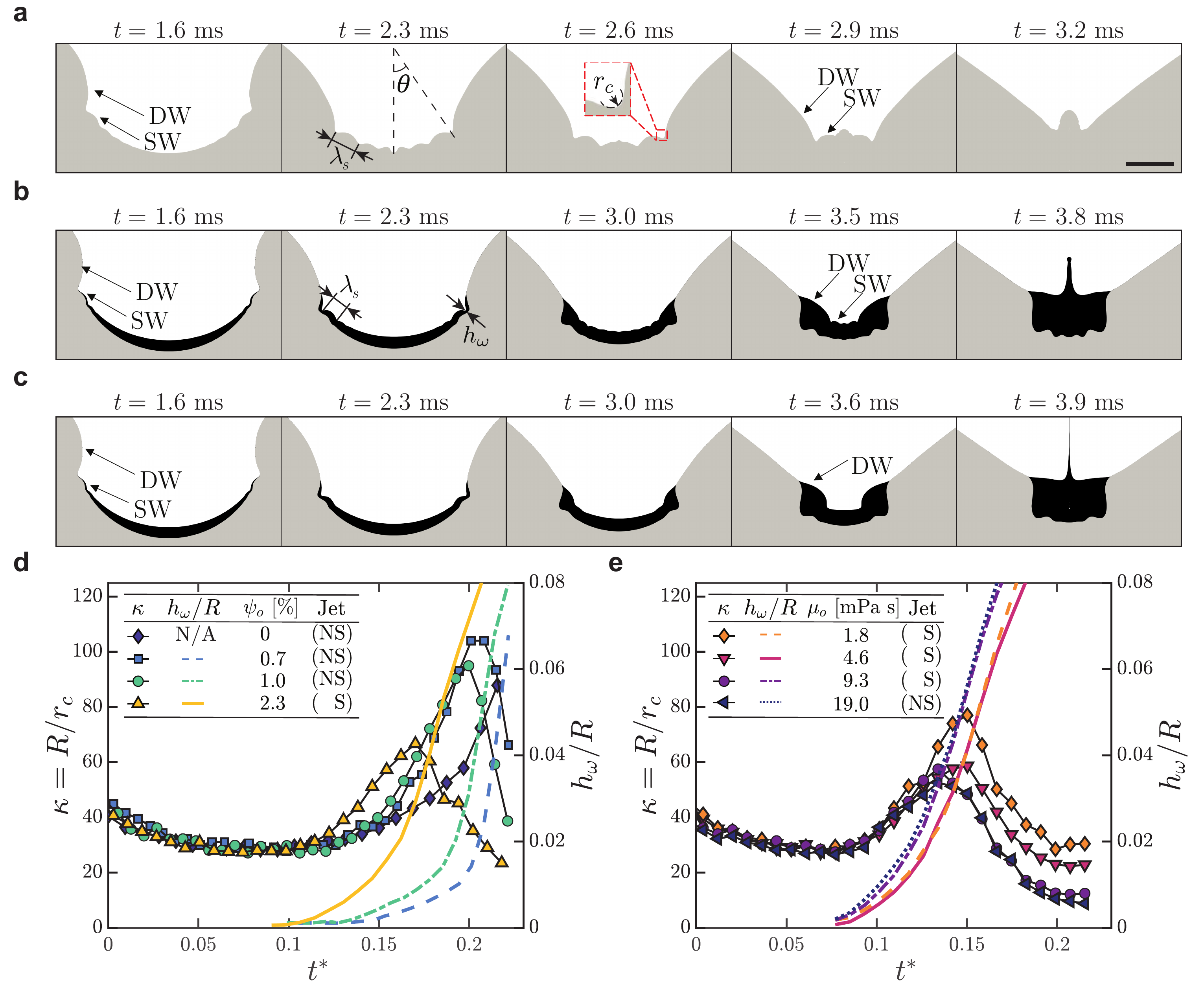}
    \caption{\textbf{Cavity collapse and capillary wave propagation.} \textbf{a-c}, Capillary wave propagation during cavity collapse after the bursting of a bare bubble (\textbf{a}), an oil-coated bubble with $\mu_o =$ 0.9 mPa s and $\psi_o = 4.2\%$ (\textbf{b}), and an oil-coated bubble with $\mu_o =$ 4.6 mPa s and $\psi_o=4.2\%$ (\textbf{c}). Inset of \textbf{a} shows the minimum radius $r_c$ measured at the dominant wave (DW) trough. The DW excites precursor waves which include the secondary wave (SW) of wavelength $\lambda_s$. The bubble radius $R=2$ mm. The scale bar represents 1 mm. \textbf{d}, Dimensionless maximum curvature $\kappa=R/r_c$ and dimensionless oil thickness at the DW trough $h_\omega/R$ as a function of dimensionless time $t^*$ at different $\psi_o$ with $\mu_o=$ 4.6 mPa s. Here,  $t^*=(t-t_{\pi/2})/t_c$, where $t_{\pi/2}$ is the time when $\theta=\pi/2$ and $t_c =\sqrt{\rho_wR^3/\gamma_e}$ is the characteristic inertia-capillary time for cavity collapse. S and NS indicate singular and non-singular jets, respectively.  \textbf{e}, $\kappa$ and $h_\omega/R$ as a function of $t^*$ at different $\mu_o$ with $\psi_o=4.2\%$. }
    \label{Sim}
\end{figure*}

To gain a quantitative understanding of how the oil coating facilitates the formation of jet singularities for the case of coated bubble bursting, numerical simulations are performed using the open-source software Basilisk\citep{basilisk} (Methods). As in previous studies\citep{deike2018dynamics,  berny2022size, sanjay2021}, the initial condition we use is given by the static shape of an oil-coated bubble obtained by solving the Young-Laplace equation\citep{lhuissier2012bursting} (Methods and Extended Data Fig.~\ref{Num_Comp}). The simulations capture the evolution of the bubble cavity, the accumulation of oil, as well as the ejected jet morphology reasonably well (Extended Data Fig.~\ref{Comparison}). Thus, these results provide detailed information of the cavity collapse and subsequent jet formation. In prior work, the liquid viscosity was found to affect the jet dynamics in two ways: (1) through damping of the precursor capillary waves that merge at the bubble base (low $Oh_o$), as well as (2) through direct damping of the jet evolution (high $Oh_o$)\cite{krishnan2017scaling,gordillo2019capillary}. Both of these effects are considered below to explain the singular regime boundary.

 After the bubble ruptures, a train of capillary waves is excited and propagates on the collapsing cavity surface as shown in Fig.~\ref{Sim}a-c, with the last wave being the most energetic\citep{ganan2021physics}. This is denoted as the ``dominant wave" (DW). Meanwhile, the precursor waves in front of the DW travel faster with shorter wavelengths. We define the one with the longest wavelength closest to the DW as the ``secondary wave" (SW), which can be clearly identified in both experiments and simulations. We observe that strong precursor wave damping is closely related to singular jet formation. For all singular cases, the precursor waves ahead of the DW are completely damped by the action of viscosity before reaching the cavity bottom pole, so that the self-similar collapse of the DW can continue closer to the singular limit unaffected by short-wavelength capillary ripples. Otherwise, the precursor waves produce strong perturbations that interfere with the self-similar collapse, preventing the formation of a singular jet. While the effect of viscosity on jet singularity has been explored in bare bubble bursting \citep{ krishnan2017scaling, brasz2018minimum}, a more sophisticated rationalization is required for the case of oil-coated bubble bursting due to the compound air-oil-water interface. 

We propose that the occurrence of the singular jet requires that the all precursor waves at the air-oil interface, including the SW, are sufficiently damped for the DW to maintain the self-similar focusing, before the viscous effect directly limits the jet velocity after the jet formation. Furthermore, the SW can serve as an indicator of this transition because it has the largest wavelength which corresponds to the smallest damping rate in all precursor waves \citep{krishnan2017scaling}. The strength of the SW is measured from the simulations using the dimensionless maximum principal curvature $\kappa=R/r_c$ ahead of the DW (inset of Fig.~\ref{Sim}a) \citep{gordillo2019capillary, blanco2021jets, sanjay2021}, where $r_c$ is the minimum radius of curvature. A smaller $\kappa$ indicates a weaker capillary wave. We observe that $\kappa$ decreases as $\mu_o$ or $\psi_o$ increases, resulting from stronger viscous dissipation effects (Fig.~\ref{Sim}d-e). Unlike in the case of bare bubble bursting, a further decrease of $\kappa$ is observed in compound bubble bursting when the local oil thickness $h_\omega/R$ exceeds a value of approximately 0.04 (Fig.~\ref{Sim}d-e). The enhanced damping effects can be interpreted as the SW fully propagating into the air-oil interface and thus experiencing a more viscous oil layer. Here, we consider $h_\omega/R$ since the wavelength of the capillary waves generated by bubble cavity collapse is found to scale with the bubble radius $R$\citep{gordillo2019capillary,sanjay2021}. For all singular cases, $\kappa$ eventually reaches a value of $\approx10-30$, consistent with the prior observation on SW for singular jet formation from bare bubble bursting \citep{gordillo2019capillary}. 


\begin{figure*}
    \centering
    \includegraphics[width=0.85\linewidth]{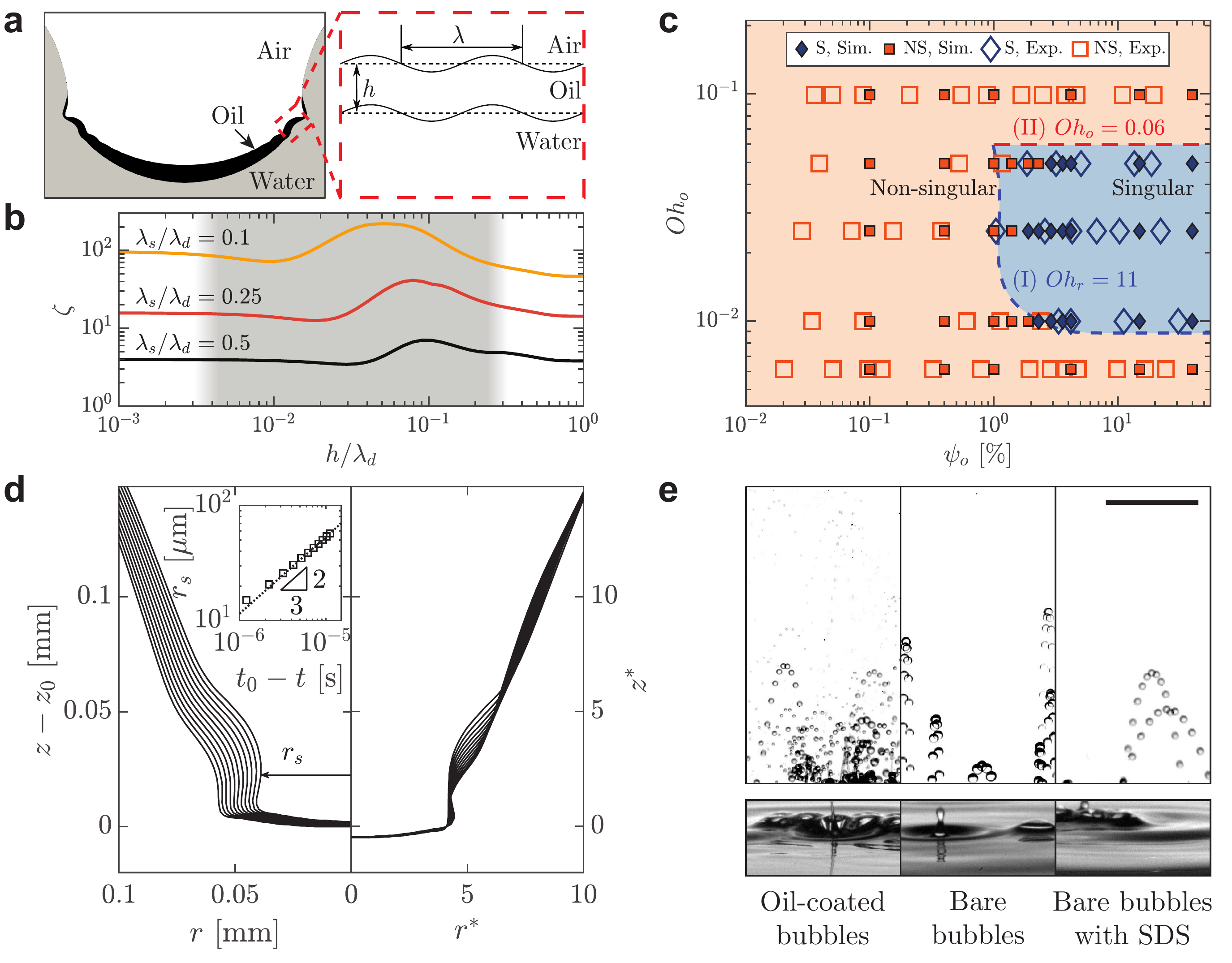}
    \caption{\textbf{Regime map of singular jets and jet drop generation by collective oil-coated bubble bursting.} \textbf{a}, Schematics of precursor waves during bubble bursting. Inset shows the model configuration where a capillary wave with wavelength $\lambda$ at a compound interface where the top liquid has a uniform thickness of $h$. \textbf{b}, Variation of dimensionless damping rate ratio of SW to DW $\zeta= T_{\lambda_s}^{-1}/T_{\lambda_d}^{-1}$ with dimensionless oil layer thickness $h/\lambda_d$ calculated by the proposed model. Here, $\mu_o/\mu_w$= 5 and $\lambda_d = R =$ \SI{2}{\mm}. The shaded area represents the range of $h$ observed in the oil-coated bubble experiments and simulations. \textbf{c}, Regime map of singular jets in both experiments and simulations. The bounding criteria are (I) $Oh_r=11$ (see Methods) and (II) $Oh_o=0.06$. \textbf{d}, Comparison between successive cavity profiles unrescaled (left) and rescaled (right) with $(t_0-t)^{2/3}$ when $\mu_o =1.8$ mPa s and $\psi_o=4.2\%$. Here,  $r^*=r(\gamma_{oa}/\rho_o)^{-1/3}(t_0-t)^{-2/3}$ and $z^*=(z-z_0)(\gamma_{oa}/\rho_o)^{-1/3}(t_0-t)^{-2/3}$, where $t_0$ represents the moment when the jet forms, and $z_0$ represents the bottom location of the entrapped bubble at $t_0$. Inset shows the minimum radius of the cavity $r_s$ versus time before the jet emerges follows a power law of 2/3. \textbf{e}, Time-lapsed images of drop ejection (top) from collective bursting (bottom) of oil-coated bubbles in pure water ($\mu_o=4.6$ mPa s, left), bare bubbles in pure water (middle), and bare bubbles in 0.8 mM sodium dodecyl sulfate (SDS) solution (right, to mimic the natural environment enriched with surface-active compounds). The bubbles are generated with a frequency of 2 s$^{-1}$. The scale bar represents 10 mm.}
    \label{Mod}
\end{figure*}

To quantitatively describe the prerequisite of the singular jet, we further evaluate the viscous damping rate of the DW and SW during cavity collapse. The viscous damping rate for the amplitude of a capillary wave with wavelength $\lambda$ at a free liquid surface can be estimated as\citep{lamb1924hydrodynamics}
\begin{equation}\label{Ts-1text}
    T_{\lambda}^{-1} = \frac{8\pi^2\mu}{\rho\lambda^2}.
\end{equation}
For bare bubble bursting, the DW and SW have wavelengths of $\lambda_d\approx R$ and $\lambda_s\approx 0.25R$, respectively\citep{ganan2017revision, krishnan2017scaling}. Therefore, the damping rates of the SW (i.e., $T_{\lambda_s}^{-1}$) and DW (i.e., $T_{\lambda_d}^{-1}$) during bare bubble bursting are correlated by a constant ratio $\zeta = T_{\lambda_s}^{-1}/T_{\lambda_d}^{-1}= (\lambda_{s}/\lambda_{d})^{-2} \approx 16$. However, for oil-coated bubble bursting, we show that $\zeta$ can be substantially enlarged due to the compound interface. To gain insight into the capillary wave dynamics in the complex, non-uniform oil layer around the collapsing cavity, we consider a simplified  set-up consisting of a capillary wave propagating at a free aqueous surface covered by a uniform oil layer of thickness $h$. We further develop a wave damping model based on the linear capillary wave theory to calculate the $\zeta$ in this configuration (see Fig. \ref{Mod}a and Methods). As shown in Fig.~\ref{Mod}b, $\zeta$ approaches $\approx (\lambda_{s}/\lambda_{d})^{-2}$ when $h$ approaches $0$ or $\infty$, with a maximum located at $h\sim\lambda_s$ (see SI section S4). This non-monotonic behavior shows that a more viscous oil layer coating the bubble cavity with $h\sim\lambda_s$, corresponding to the current experiments, may significantly increase the ratio of the damping rates between the SW and DW relative to that at a clean interface. Therefore, the compound interface favors the production of a singular jet by smoothing out the shorter wavelength perturbations.

In addition, in the case of oil-coated bubble bursting, the capillary waves encounter an oil layer with non-uniform thickness, which leads to a capillary wave separation, thus further increasing the damping rate ratio between SW and DW. This can be seen by considering the wave speed and wavelength of SW during the oil-coated cavity collapse. The wave speed is set during the initial film rupture at the top air-oil-water interface, given by $U \sim v_{ce}$, as confirmed by the experiments and simulations (Extended Data Fig. \ref{fig:WaveProp}c). As the SW propagates, it encounters an oil layer of increasing thickness and splits between the air-oil and oil-water interfaces while maintaining a nearly constant wave speed. As the oil layer thickness increases, the SW begins to experience a different bulk liquid with a different surface tension, resulting in a shorter $\lambda_s$ than the case without the wave separation (Extended Data Fig. \ref{fig:WaveProp}d). Thus, the presence of an oil layer decreases $\lambda_s/\lambda_d$ (from $\approx$ 0.25 to 0.1, see Methods), which further increases $\zeta$ by more than one order of magnitude (up to $\approx$~220) compared to bare bubble bursting (Fig.~\ref{Mod}b). The significant increase of damping rate thus facilitates the formation of a singular jet over a wider parameter space by relatively increasing the damping of the SW, allowing the DW to experience the self-similar collapse.

Based on our modeling of capillary wave damping, we now rationalize the bounding criteria for singular jets from compound bubble bursting. For bare bubble bursting, $Oh$ can be interpreted as $Oh\sim T_{\lambda_s}^{-1}/t_{c}^{-1}$, the ratio between the damping rate of the SW and the inverse of the inertia-capillary timescale (i.e. $t_{c}^{-1}=(\rho R^3/\gamma)^{-1/2}$). However, for compound bubble bursting, we obtain $T_{\lambda_s}^{-1}$ with our model calculation for SW damping at different $\psi_o$ and $Oh_o$, and we propose a revised Ohnesorge number $Oh_r$ in place of $Oh$ as 
\begin{equation}\label{Fetext}
   Oh_r=\frac{T_{\lambda_s}^{-1}}{t_{bc}^{-1}},
\end{equation}
where $t_{bc}\approx0.3\sqrt{\rho_w R^3/\gamma_e}$ for oil-coated bubbles obtained from our experiments and simulations. We find that the isoline of $Oh_r = 11$ aligns well with the lower and left boundaries of the singular jet regime (Fig. \ref{Mod}c), which is quantitatively analogous with the lower critical $Oh\approx0.03$ for singular jets from bare bubble bursting in a single liquid\citep{duchemin2002jet, deike2018dynamics, brasz2018minimum, blanco2020sea} (Methods and SI section S5). Therefore, the SW damping is responsible for setting the boundary (I) ($\psi>1\%$ and $Oh_o>0.01$) in Fig.~\ref{Mod}c, which is captured by our proposed $Oh_r$. In addition, the numerical results for singular jetting confirm that the minimum cavity radius follows the  inertia-capillary self-similarity behavior with the power law $r_s\sim (t_0-t)^{2/3}$, where $t_0$ indicates the singular time when the cavity inverts (Fig.~\ref{Mod}d). 

Furthermore, with a further increase of $Oh_o$, the viscous stresses continuously dampen the DW, limiting the jet velocity and enlarging the top jet drop due to the delay of jet breakup. This excess viscous damping results in a maximum $Oh_o$ for which singular jetting can occur. This transition is shown as the boundary (II) in Fig.~\ref{Mod}c and corresponds to $Oh_o\approx0.06$, consistent with the transitional value of $Oh$ in bare bubble bursting when the viscous stresses become strong enough to directly suppress cavity cusp formation\citep{lee2011size,brasz2018minimum, blanco2020sea,raja2020conditions}. In addition, when the bulk viscosity $\mu_w$ varies from 1-22.5 mPa s corresponding to $Oh_w=\mu_w/\sqrt{\rho_w\gamma_{e}R}$ of $O(10^{-3}-10^{-1})$, singular jetting occurs when $\psi_o > 1\%$ (Extended Data Fig. \ref{Bulk}), while bare bubble bursting only produces singular jets within a narrow range of $Oh$ ($\approx 0.02-0.05$)\citep{brasz2018minimum,gordillo2019capillary} for the bulk liquid. These results show that the compound interface with the oil coating could facilitate the inertia-capillary self-similarity, expanding the regime of singular jetting in bubble bursting and decreasing the jetted drop sizes.

More generally, our study on oil-coated bubble bursting demonstrates the hitherto unrecognized role of the compound interface on the bubble-driven aerosol flux. In particular, due to the wider parameter space for singular jetting, collective oil-coated bubble bursting tends to generate jet drops with smaller sizes, overall greater numbers of drops, and higher droplet ejection heights compared with bare bubble bursting at either clean or surfactant-laden aqueous surfaces as shown in Fig.~\ref{Mod}e. Here, a sodium dodecyl sulfate solution was used to mimic a natural water body enriched by surface-active compounds.
 The droplet size is one key parameter in predicting its residence time and transport, since small droplets are more easily lifted by turbulent eddies\citep{veron2015ocean}. In addition, these contaminant-laden drops smaller than 10 $\mu$m in diameter may pose a higher risk of pollutant spread or infection since they can penetrate further into the respiratory tract than larger drops\citep{gralton2011role, bourouiba2020}. The oil-coated bubbles in our experiments could typify the ubiquitous contaminated or compound bubbles in the oceans, and the bubble-bursting jet drop particles have been found to contain a different composition with stronger ice nucleating ability than film drop particles\citep{wang2017role}. Hence, our discovery may potentially improve chemical transport modeling related to bubble-driven flux regarding sea spray aerosols. In industry, these small drops resulting from the singular jets produced by compound bubble bursting may impose detrimental impacts to the workers' health, such as the generation of acidic mist in electrolysis 
 \citep{ma2020characteristics} and bioaerosols from wastewater treatment plants\citep{ginn2021detection, lou2021bioaerosols}. Our work may suggest additional guidelines for personal protective equipment and management controls on air and water quality near these facilities \citep{law2021covid}. Meanwhile, bubble bursting is considered as the major cause for cell damage in bioreactors via the hydrodynamic stresses produced by cavity collapse and jet breakup \cite{mcrae2021aerosol}. The thin and fast singular jet regime from compound bubbles may sharply increase the stresses, and thus require a more careful control of aeration and agitation. In closing, these results on the production of singular jets by oil-coated bubble bursting offer new insights into the dynamic processes of complex fluids, with potential environmental consequences and industrial applications.

\newpage
\clearpage
\noindent \textbf {METHODS}
\bigbreak
\noindent \textbf {Materials} 

\noindent Deionized water (resistivity = 18.2 M$\rm \Omega\cdot$cm) was obtained from a laboratory water purification system (Smart2Pure 3 UV/UF, Thermo Fisher Scientific). Octamethyltrisiloxane (referred as the silicone oil with kinematic viscosity $\nu = $ 1 cSt), dodecamethylpentasiloxane (silicone oil with $\nu = $ 2 cSt), other silicone oils ($\nu$ = 5, 10, 20 cSt), and sodium dodecyl sulfate (BioXtra, $\ge99.0\%$ were purchased from Sigma-Aldrich and used as received. Glycerin was purchased from Fisher Chemical. The surface tensions $\gamma_{oa}$ (or $\gamma_{wa}$) of the liquids and the interfacial tensions between silicone oils and water $\gamma_{ow}$ were measured using the pendant drop method, and the densities $\rho$ and dynamic viscosities $\mu$ are listed in Extended Data Table 1.
\bigbreak
\noindent \textbf {Experimental setup} 

\noindent The experimental apparatus is shown in Extended Data Fig. \ref{ExpSche}. A square transparent acrylic container of $20\times20\times25$ mm$^3$ was fabricated to hold the liquids, and we measured the contact angle of water on the acrylic to be $86\pm8^\circ$. We used the custom-designed coaxial orifice system detailed in our previous work \citep{ji2021oil,ji2021oilcolumn} to produce oil-coated bubbles. For the coaxial orifice system, the inner diameter of the inner needle was $d_{ni} = $ \SI{0.41}{\mm}, and the outer diameter of the outer needle was $d_{po} = $ \SI{3.43}{\mm}. The equilibrium radius of the compound bubble (gas+oil) in our experiments was determined to be $R \approx 2$ mm. 

Two high-speed cameras (FASTCAM Mini AX200, Photron) were used to synchronously record the top and bottom side views of the oil-coated bubble bursting at a free liquid surface, separately illuminated by two LED panels. We carefully maintained a slightly convex meniscus at the air-water surface over the container edge by filling up the container, which prevented the bubbles from drifting to the side of the container out of the focal plane \citep{neel2021collective}. This method also allowed better imaging of the jet with the meniscus slightly lower than the free surface. In addition, by tilting the cameras with an angle of $\approx 5^\circ$, the influence of the meniscus on the visualization can be further avoided. We used a frame rate of 6400-20000 frames per second and a magnification of 1-4. We also used an advanced high speed camera (FASTCAM SA-Z, Photron), with a frame rate of 50000 frames per second and a magnification of 12.3 to obtain high resolution images. The obtained images were post-processed with Fiji and MATLAB. The volume of the oil in the oil-coated bubble $V_o$ was estimated by measuring the oil volume at the bubble bottom in the high-speed video before bubble bursting, and then the oil fraction was calculated as $\psi_o=3V_o/\left(4\pi R^3\right)$. 

For the collective bubble bursting (Fig.~\ref{Mod}e), the bare gas bubbles were generated with a needle of a diameter = 3.43 mm. The equilibrium gas bubble radius was determined to be $2.3\pm0.2$~mm, similar to the oil-coated bubble radius. In each experiment, we took a high-speed video with a duration of 11 s at a frame rate of 125 frames per second. All top-view images were overlapped together to produce the upper row of Fig.~\ref{Mod}e.

To provide more information of the jet drop composition, a cobalt chloride test strip for water detection (PGA01V100, Bartovation, NY, US) was used to collect the jet drops by bubble bursting to test the presence of water. If the drop contact location turns pink, the jet drop contains water \citep{kan2021humidity}. We note that we performed control experiments with micropipette tips to manipulate the deposited drop size, and found that the color change is observable for a water drop with a radius as small as \SI{15} {\micro\meter}. In our bubble bursting experiments with 4.6 mPa s oil, when $\psi_o=0.5\%$, the jet drop radius was larger than \SI{100}{\micro\meter}. There was no color change of the test strip already. Thus, the jet drops should only contain oil for $\psi_o\ge 0.5\%$ since our control experiments show that water can be detected in drops with a radius of \SI{15} {\micro\meter}. This critical $\psi_o$ is further confirmed by our numerical simulation with 4.6 mPa s oil (Extended Data Fig.~\ref{fig:droptest}), while the oil volume composition of the top jet drop is $\approx$ 10\%-40\% when $\psi_o<0.5\%$ and 100\% when $\psi_o\ge0.5\%$.


\bigbreak
\noindent \textbf {Numerical simulations} 

\noindent Numerical simulations were performed using the open-source partial differential equation solver Basilisk\citep{popinet2015quadtree,beetham2016wave,popinet2003gerris}. The Basilisk  solver is especially well-suited to performing adaptive mesh refinement, which is critical for resolving such multiphase flow problems with jetting\citep{deike2018dynamics}, droplet breakup\citep{brasz2018threshold}, drop impact and splash\citep{marcotte2019ejecta,fudge2021dipping} and thin films\citep{agbaglah2021breakup}. In particular, axisymmetric simulations were performed using the three-phase solver developed by Sanjay et al. \citep{sanjay2022taylor}. 

For the simulations, we set the initial conditions as an oil-coated bubble resting at an air-liquid interface with a hole to connect the bubble interior to the gas phase above the interface, which represents a symmetrically rupturing bubble cap. We have developed a model to calculate the initial static oil-coated bubble shape (see Methods section `Initial static bubble shape' and SI section S2), which precisely reproduces the experimental static bubble shape as shown in Extended Fig. \ref{Num_Comp}. In each case, the oil-coated bubble was initialized in a 15 mm square domain where $r$ ranges from 0 to 15 mm and $z$ ranges from -7.5 to 7.5 mm. The $z$ origin of the bubbles is shifted to give sufficient room to resolve the jetting drops and such that the water at the bottom of the domain is relatively undisturbed. We use a minimum refinement level of 5, corresponding to a maximum simulation cell size of $\sim0.469$ mm and a maximum refinement level of 13, corresponding to a minimum simulation cell size of $\sim1.83$ $\mu$m. The initial condition is always resolved up to the maximum refinement level, and then the adaptive mesh refinement takes over, always resolving any fluid interfaces to the maximum level. Specifically, the \texttt{adapt} function is used to control the adaptive meshing each time step with tolerance values on the interface volume fraction fields, the velocity components, and the curvatures of the air-oil and oil-water interfaces of $1\times10^{-3}$ for each tolerance. The imposed non-default boundary conditions in the simulations were a zero normal gradient condition on the velocity and a fixed zero Dirichlet boundary condition at the top boundary. The solver automatically controls the time step to guarantee stability based on the surface tensions, velocities, and adaptive meshing, and output data files were written every 0.1 or 0.01 ms. A pre-wetting oil film with a layer of exactly one cell thickness is assumed at the air-water interface\citep{sanjay2022taylor,afkhami2018transition}. For the configuration of oil-coated bubble bursting in water, such a numerical assumption with a pre-wetting film is applicable only when it is thermodynamically favorable for oil to spread on water, i.e. the spreading coefficient 
$S=\gamma_{aw}-\gamma_{oa}-\gamma_{ow}\ge 0$\citep{bonn2009wetting,de2004capillarity, sanjay2022taylor}, as in our current experiments with silicone oil and water. A typical runtime of $\sim 500$ CPU hours is used for each of these cases, and we run each simulation in parallel on 32 processors.  

The numerical simulations were validated by comparing with experiments of bare bubble bursting at pure liquid surfaces regarding the jet tip radius and velocity. For the oil-coated bubble, our simulation shows good agreements with experimental results with respect to the time for cavity collapse and overall shape of the interfaces (Extended Data Fig.~\ref{Comparison}). We also performed simulations with different refinement levels as convergence tests to show that our results are independent of the mesh refinement level. With the current refinement level, the non-singular jet velocity and radius are already converged, \textit{viz}.\ they remain unchanged when the refinement level is increased. However, the singular jet velocity and radius will not converge as the refinement level is increased. Therefore, we consider the convergence tests as distinguishing between regimes that illustrate singular jet formation and the regimes that do not. To that end, we have performed simulations with maximum refinement levels of 11, 12, and 13 and found that they all predict the same singular jet regimes (see detailed discussion in SI section S6). In that sense, and for the purpose of the current work, we consider these results converged, and we opted for the higher refinement level for the best accuracy possible. We note that for the bare bubble bursting, such a maximum level of 13 in the simulation with Basilisk has been adopted by prior work for bubble-bursting jets\citep{lai2018bubble,deike2018dynamics}. In addition, our simulations assume an axisymmetric flow as most prior simulation work does\citep{brasz2018minimum,gordillo2019capillary}, but bubble bursting in reality might show asymmetry from film rupture to the final jet formation and breakup\citep{lhuissier2012bursting}. Furthermore, it has been suggested that simulation prediction for the singular jet velocity and radius appears difficult to compare with experiments due to the convergence issues\citep{mou2021singular}. All the above factors may contribute to the discrepancies between the experiments and simulations, in particular regarding the singular jet tip velocity and radius.

Within the scope of the current work, we observe strong agreement comparing the numerically and experimentally determined parameter regimes for singular jet formation (Fig.~\ref{Mod}c). In addition, regarding the SW propagation contributing to the jet formation, we also obtain a great consistency as shown in Extended Data Fig.~\ref{fig:WaveProp}. Therefore, we believe our simulations provide reliable and insightful information to understand the jet formation in oil-coated bubble bursting.

\bigskip
\noindent \textbf {Initial static bubble shape} 

\noindent The initial static shape of an oil-coated bubble resting at the water surface is calculated by numerical solutions of the Young-Laplace equation. The assumed static bubble shape is separated into several interface portions and solved iteratively with matching boundary conditions, following a similar approach as the prior work \citep{ lhuissier2012bursting} (See SI section S2).

\bigskip
\noindent \textbf {Capillary wave damping model} 

\noindent To rationalize the damping on the capillary waves during the oil-coated cavity collapse, we present a simplified model to describe the effect of the oil coating on cavity collapse and jet formation. As shown in Fig. \ref{Mod}a, a layer of oil with a uniform thickness $h$ at rest is deposited on a deep bath of water. Here, we neglect the bubble cavity curvature and film thickness variation considering the fact that $h_{max}/R\ll1$ in the experiments. In addition, the gravitational effects are negligible given that the typical capillary wavelength in bubble bursting is smaller than the capillary length $\sqrt{\rho_{oa}/(\rho_og)}$.  We consider a periodic travelling capillary wave with the long-wave approximation, which has been widely used to model the waves excited by bubble cavity collapse\citep{zhang2015partial, gordillo2019capillary}. In addition, we assume that non-linear effects are negligible in order to use the linearized Navier-Stokes equations. This can be justified because the boundary layer that develops around the wave crest has a thickness $\delta \approx \sqrt{2\mu_o/(\rho_o\omega)}$, where the angular frequency $\omega$ can be estimated as $\sqrt{(\gamma_{oa}/\rho_o)(2\pi/\lambda)^3}$\citep{gordillo2019capillary,longuet1992capillary}. In all our experiments, $\delta/\lambda<0.2$.
Following the linear capillary wave theory \citep{lamb1924hydrodynamics, landau2013fluid, levich1962, whitham1974linear, jenkins1997wave}, we derive the dispersion relation using the linearized Navier-Stokes equation (see SI section S4). By numerically solving the dispersion relation, we obtain the wave damping rate, which is further validated with the prior theoretical results\citep{lamb1924hydrodynamics} in the limit of $h \rightarrow 0$ (see SI section S4). We use our derivation to estimate the damping rate of the SW and DW for the oil-coated cavity collapse and calculate $Oh_r$.

Notably, our linear wave damping model has limitations with respect to short capillary waves where the long-wave approximation might not hold, or for highly viscous liquids where the boundary layer thickness $\delta\sim\lambda$. In the latter case, the viscous damping in the rotational flow within the boundary layer needs consideration, and thus non-linear effects cannot be neglected \citep{denner2016frequency,gordillo2019capillary}.

\bigskip
\noindent \textbf {Capillary wave separation}

\noindent Here, we focus on further characterization of the capillary wave propagation during cavity collapse since these waves directly dictate the formation of the jet. Extended Data Fig. \ref{fig:WaveProp}a-b shows the propagation of capillary waves in the case of a compound or bare bubble bursting and the SW is highlighted. While the wave speed remains almost the same, the wavelengths and amplitudes of the SW in the compound bubble case are smaller. The capillary wave speed $U$ and wavelength $\lambda$ are associated with the dispersion relationship\citep{lamb1924hydrodynamics} as
\begin{equation}
        \lambda = \frac{2\pi\gamma}{\rho U^2}. \label{dispersion}
\end{equation}
which has been applied to analyze the capillary wave propagation resulting from bubble cavity collapse with a small $Oh$\citep{zhang2015partial, krishnan2017scaling}. The variation of the angular position of the DW trough $\theta$ over time has been used to characterize the wave propagation speed as $U=R \md\theta/\md t$.\citep{gordillo2019capillary}

For bare bubble bursting in pure liquids, ${U/v_{cw}}\approx 5$, where the characteristic capillary velocity for a clean bubble is $v_{cw}= \sqrt{\gamma_{wa}/(\rho_w R)}$\citep{gordillo2019capillary,blanco2020sea,blanco2021jets}. Thus, the wavelength of the SW for bare bubble bursting, $\lambda_{sp}$, is obtained as
\begin{equation}
    \frac{\lambda_{sp}}{R} \approx \frac{2\pi}{25}. \label{lP}
\end{equation}
We note that this prediction is consistent with the previous work\citep{krishnan2017scaling,blanco2020sea}. Regarding oil-coated bubbles with a compound interface, the capillary waves split into the air-oil and oil-water interfaces when the oil layer becomes thick. However, during the whole cavity collapse, ${U}/v_{ce}$ is also found to be $5.7 \pm 0.7$ in both our experiments and simulations, as shown in Extended Data Fig. \ref{fig:WaveProp}c. When the SW has fully entered the air-oil interface after the wave separation, the oil density $\rho_o$ and surface tension $\gamma_{oa}$ are to be considered in equation (\ref{dispersion}), so we obtain  
\begin{equation} \label{lambdac}
    \frac{\lambda_{s, ao}}{R} \approx  \frac{\rho_w}{\rho_o}\frac{\gamma_{oa}}{\gamma_{e}}\frac{2\pi}{25}\approx \frac{\rho_w}{\rho_o}\frac{\gamma_{oa}}{\gamma_{e}}\frac{\lambda_{sp}}{R}.
\end{equation}

With simulations, we show that the dimensionless SW wavelength of compound bubble bursting at $\theta=\pi/6$ decreases significantly as $\lambda_{s,ao}/R=0.11$ after capillary wave separation for large $\psi_o$ compared to that for a bare bubble, which is $\lambda_{sp}/R = 0.28$ (Fig.~\ref{fig:WaveProp}d). The measurements agree well with our prediction in \cref{lP,lambdac}. Our results confirm that the compound interface also contributes to the decrease of the SW wavelength, and thus increases the corresponding damping rate.

\bigskip
\noindent \textbf {Nondimensional number $Oh_r$ for singular jets}

\noindent It has been shown that the decrease of capillary wave amplitude during cavity collapse of bare bubble bursting could be described by $Oh$ \citep{krishnan2017scaling,gordillo2019capillary}. At a free liquid surface, the amplitude $\alpha$ of capillary waves falls off exponentially in the form $\alpha=\alpha_0\mathrm{e}^{-t/T_{\lambda}}$ with the damping rate calculated by equation (\ref{Ts-1text}). In the context of bare bubble bursting in a pure liquid, Krishnan et al.\citep{krishnan2017scaling} propose that in the bubble collapse time $t_{bc}\approx 0.3 t_c$, where $t_{c}=\sqrt{\rho R^3/\gamma}$, so that the decrease in capillary wave amplitude is given by 
\begin{equation}\label{Dp}
    \ln{\left(\frac{\alpha}{\alpha_0}\right)}=-\frac{t_{bc}}{T_{\lambda}} \approx -24\left(\frac{R}{\lambda}\right)^2 Oh.
\end{equation}
They observed that the capillary waves are progressively damped as $Oh$ increases, and the singular jet occurs at a critical $Oh$ due to this damping in bare bubble bursting. Such an observation indicates that when $|\ln{\left({\alpha}/{\alpha_0}\right)}|$ increases to a critical value, all the precursor capillary waves are damped and thus the DW is more effectively focused for a self-similar collapse. Therefore, we further consider such a capillary wave damping theory with respect to $\ln{\left({\alpha}/{\alpha_0}\right)}$ for our oil-coated bubble bursting jets. We propose a non-dimensional number, $Oh_r$, defined as 
\begin{equation}\label{Fe}
    Oh_r=\left|\ln{\left(\frac{\alpha}{\alpha_0}\right)}\right|=\frac{t_{bc}}{T_{\lambda_s}},
\end{equation}
which can be interpreted as the damping of precursor capillary waves merging at the bubble base during cavity collapse. The damping rate of the SW, $T_{\lambda_s}^{-1}$, can be predicted with our proposed model (see SI Section S5). 

Since the SW has the longest wavelength among the precursor capillary waves and thus the minimum damping rate, the attenuation of all precursor capillary waves could be described solely by the non-dimensional parameter $Oh_r$. In both the experiments and simulations, we find that the oil-coated bubble collapse time $t_{bc}\approx0.3t_{co}$, where $t_{co}=\sqrt{\rho_w R^3/\gamma_e}$, and $T_{\lambda s}$ is calculated with $\lambda_s/R\approx0.1$ using our proposed model for wave damping. The limit of $Oh_r\approx11$, which indicates the same damping rate of the SW, is found to accurately describe the left and bottom boundaries of the singular regime in the $\psi_o-Oh_o$ regime map for oil-coated bubble bursting. We note that our model does not include any fitting parameters. In addition, considering $Oh_r=|\ln{\left({\alpha}/{\alpha_0}\right)}|=11$ and $\lambda_s/R\approx0.25$ in bare bubble bursting, the lower critical $Oh_c$ for singular jetting of bare bubble bursting could be calculated to be $Oh_c\approx0.03$ using equation (\ref{Dp}), which is consistent with the literature values found experimentally and numerically \cite{krishnan2017scaling,gordillo2019capillary,brasz2018minimum}.

\bigskip

\noindent \textbf {Data availibility} The data used in this study are available from the corresponding author upon reasonable request.
\bigskip

\noindent \textbf {Code availibility} The codes used for Basilisk simulation in this study are available at \url{http://basilisk.fr/sandbox/jtault/README}. The codes for bubble shape calculation are available at \url{https://github.com/zyyang-mech/Enhanced-singular-jet-formation-in-oil-coated-bubble-bursting}.

\bigskip

\noindent \textbf {Acknowledgements} We thank Professor Howard A. Stone at Princeton University for helpful discussion about the manuscript, Professor Sascha Hilgenfeldt at the University of Illinois at Urbana-Champaign for discussion on wave modelling, and Dr. Vatsal Sanjay at the University of Twente for fruitful suggestions about the simulations. This work is partially supported by the American Chemical Society Petroleum Research Fund Grant No. 61574-DNI9 (to J.F.) 

\bigskip
\noindent \textbf {Author contributions} B.J. and J.F. conceived the project. B.J. and J.F. designed the experiments. Z.Y. and B.J. conducted the experiments and analysed the results. Z.Y., B.J. and J.F. conducted the theoretical analysis. J.T.A. conducted the simulations with Basilisk. Z.Y. conducted other numerical analyses. J.T.A. and Z.Y. post-processed the simulation results. Z.Y., B.J., J.T.A. and J.F. discussed the results and wrote the manuscript.

\bigskip
\bibliographystyle{sn-aps}
\bibliography{sn-bibliography}

\newpage
\begin{exttable*}
\caption{\label{tab:example}Physical properties of the liquids used in the experiments ($\rho$: density; $\mu$: dynamic viscosity; $\gamma_{wa}$: surface tension of water or aqueous solutions; $\gamma_{oa}$: surface tension of oil; $\gamma_{ow}$: oil-water interfacial tension).}
\centering
\begin{tabular}{cccccc}
\hline
Liquids  & $\rho$ ($\rm{kg/m^3}$)  &  $\mu$ (mPa s)& $\gamma_{wa}$ (mN/m)  & $\gamma_{oa}$ (mN/m) & $\gamma_{ow}$ (mN/m) \\[3pt]
\hline
DI water   & 998 & 1.0 & $71.6\pm1.0$ & N/A & N/A\\

1 cSt silicone oil   & 820 & 0.9 & N/A &$13.1\pm0.4$ & $27.3\pm0.3$\\

2 cSt silicone oil   & 875 & 1.8 & N/A &$18.6\pm0.2$ & $35.0\pm0.2$\\

5 cSt silicone oil   & 913 & 4.6 & N/A & $18.7\pm0.3$ & $38.1\pm0.4$\\

10 cSt silicone oil   & 930 & 9.3 & N/A &$19.1\pm0.3$ & $40.9\pm0.5$\\

20 cSt silicone oil  & 950 & 19.0 & N/A & $19.4\pm0.7$ &  $40.9\pm0.5$ \\

20 wt\% glycerin-water solution & 1047 & 1.8 &$70.9\pm0.5$ &N/A & N/A\\

40 wt\% glycerin-water solution & 1099 & 3.7 &$66.9\pm0.4$ &N/A & N/A\\

50 wt\% glycerin-water solution & 1126 & 6.0 &$66.4\pm0.5$ &N/A & N/A\\

70 wt\% glycerin-water solution & 1181 & 22.5 &$61.4\pm0.3$ &N/A & N/A\\
\hline

\end{tabular}
\end{exttable*}

\begin{extfigure*}
    \centering
    \includegraphics[width=0.5\linewidth]{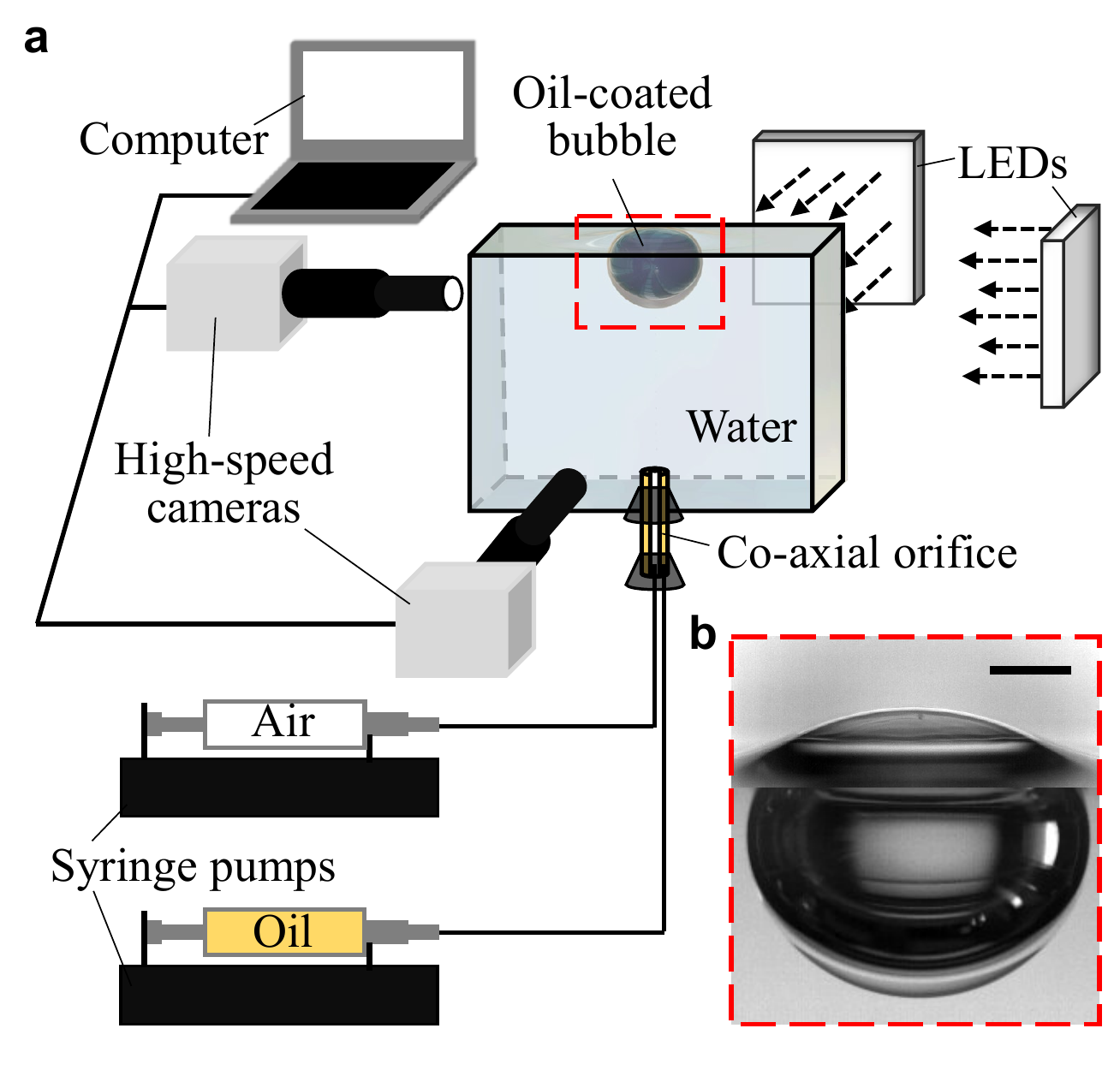}
    \caption{\textbf{Experiment setup for the imaging of oil-coated bubble bursting.} \textbf{a}, Schematic drawings of the experiment setup. The oil-coated bubbles are generated from coaxial orifices, and observed with two high-speed cameras simultaneously. \textbf{b}, Zoom-in image of a typical oil-coated bubble at a free surface with $\mu_o =19$ mPa s and $\psi_o = 6\%$.}
    \label{ExpSche}
\end{extfigure*}

\begin{extfigure*}
    \centering 
    \includegraphics[width=0.4\linewidth]{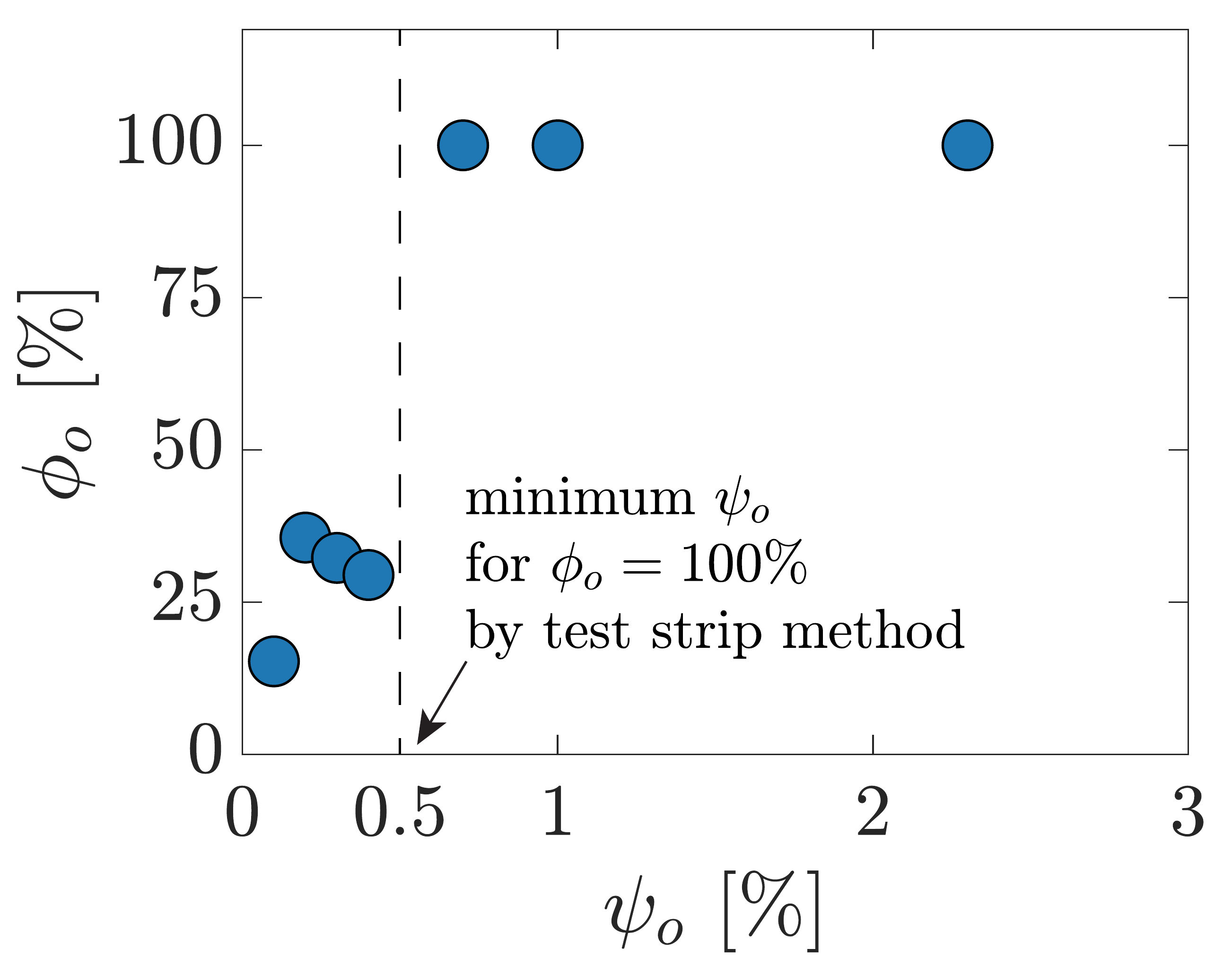}
    \caption{\textbf{Variation of oil volume composition in top jet drop from bursting of oil-coated bubbles with different oil volume fractions.} The data points denote the oil volume composition of the top jet drop for oil-coated bubble bursting (with 4.6 mPa s oil), obtained from simulations. Here $\phi_o$ represents oil volume composition in the top jet drop. The dashed line denotes the minimum oil volume fraction to produce an oil-only top jet drop estimated from experiments.}
    \label{fig:droptest}
\end{extfigure*}

\begin{extfigure*}
    \centering 
    \includegraphics[width=0.9\linewidth]{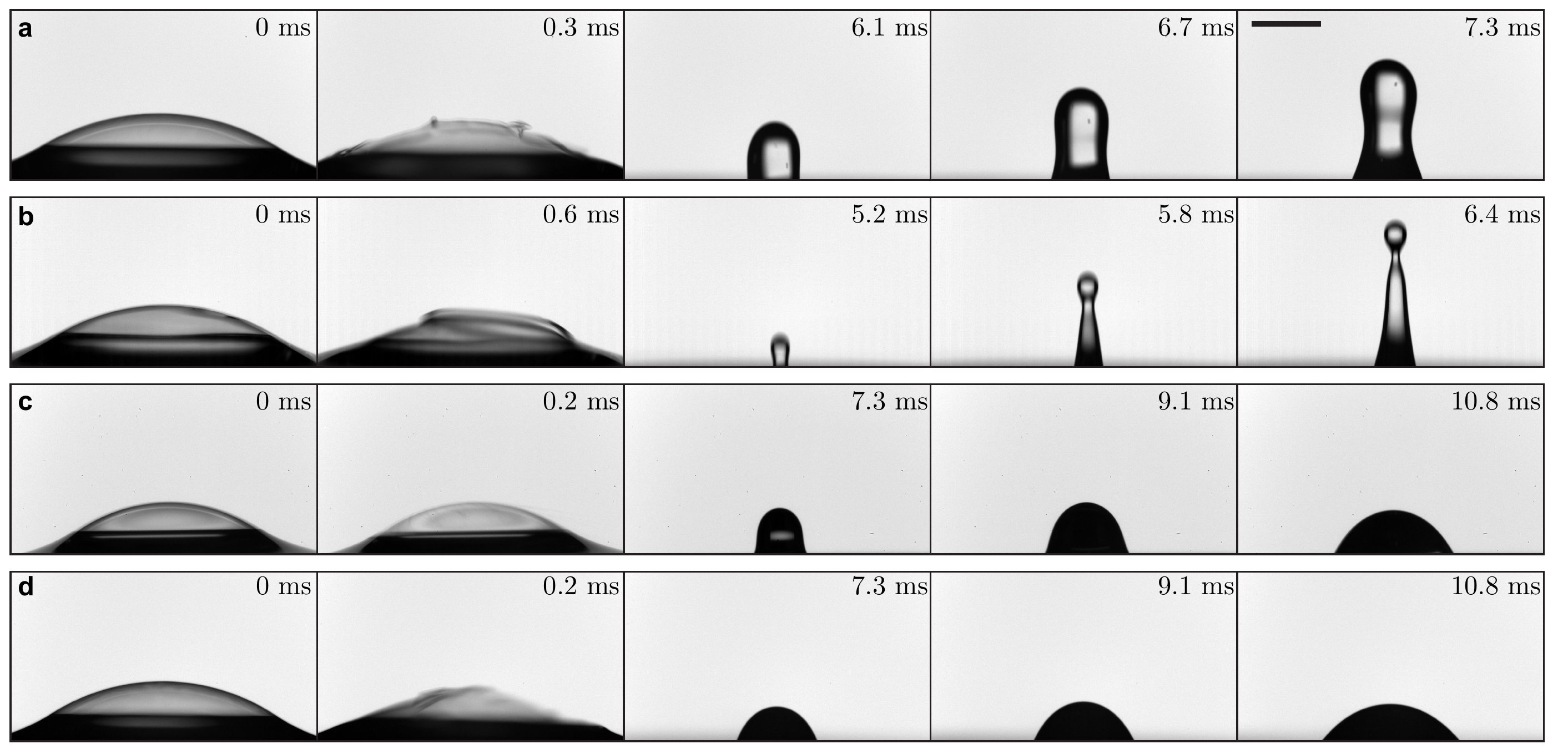}
    \caption{\textbf{Bursting of bare bubbles of $\boldsymbol{R} \approx$~2 mm at liquid surfaces with increasing $\boldsymbol{Oh_w}$ only produces non-singular jets.} A bare bubble with $R=2.1\pm 0.3$ mm bursts at the surface of water ($Oh_w=0.0026$, \textbf{a}), 50 wt\% glycerin-water solution ($Oh_w=0.015$, \textbf{b}), 4.6 mPa s silicone oil ($Oh_w=0.025$, \textbf{c}), and 70 wt\% glycerin-water solution ($Oh_w=0.06$, \textbf{d}). The scale bar represents \SI{1}{\mm}. }
    \label{fig:nonsingular}
\end{extfigure*}

\begin{extfigure*}
    \centering 
    \includegraphics[width=0.9\linewidth]{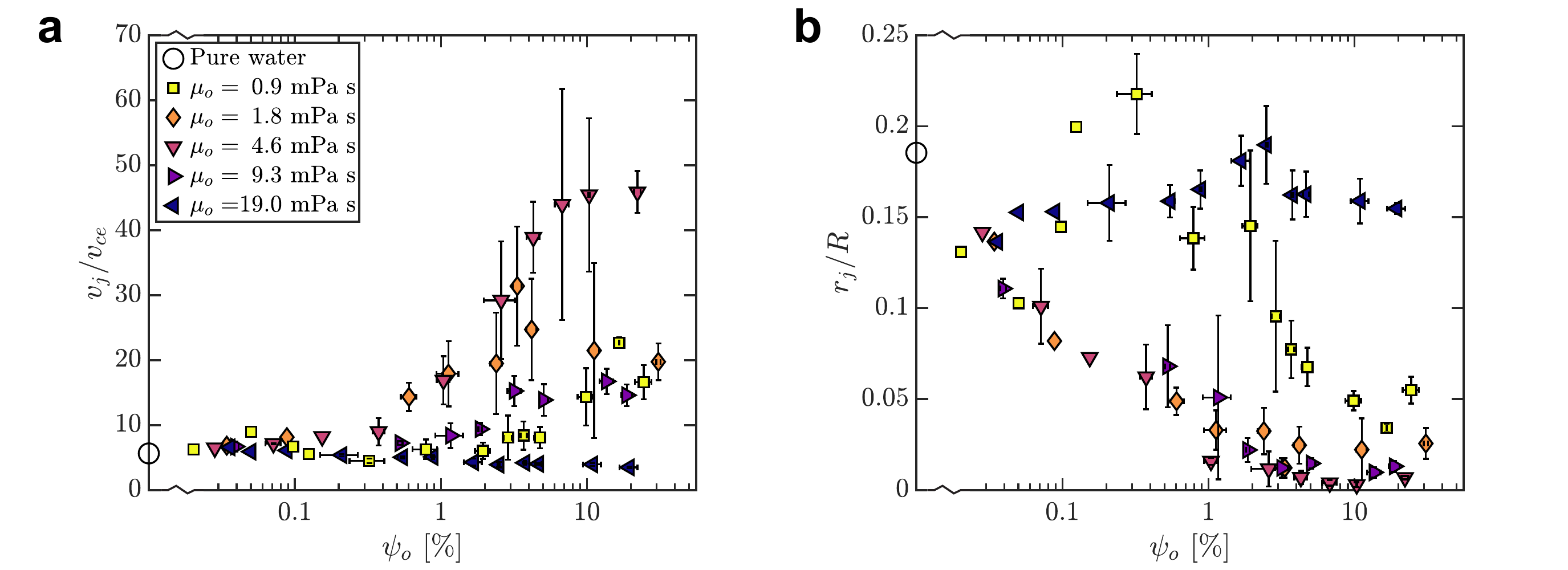}
    \caption{\textbf{Dimensionless jet velocity and radius from oil-coated bubble bursting.} \textbf{a}, Dimensionless jet velocity $v_j/v_{ce}$ as a function of oil volume fraction $\psi_o$ at different oil viscosities $\mu_o$. For the pure water case, $v_{ce} = v_{cw}$ is used. \textbf{b}, Dimensionless jet radius $r_j/R$ as a function of $\psi_o$. The hollow markers at the left vertical axes of \textbf{a}, \textbf{b} represent the case for a bare bubble of the same size bursting in pure water.}
    \label{fig:dimlessext}
\end{extfigure*}

\begin{extfigure*}
    \centering
    \includegraphics[width=0.8\linewidth]{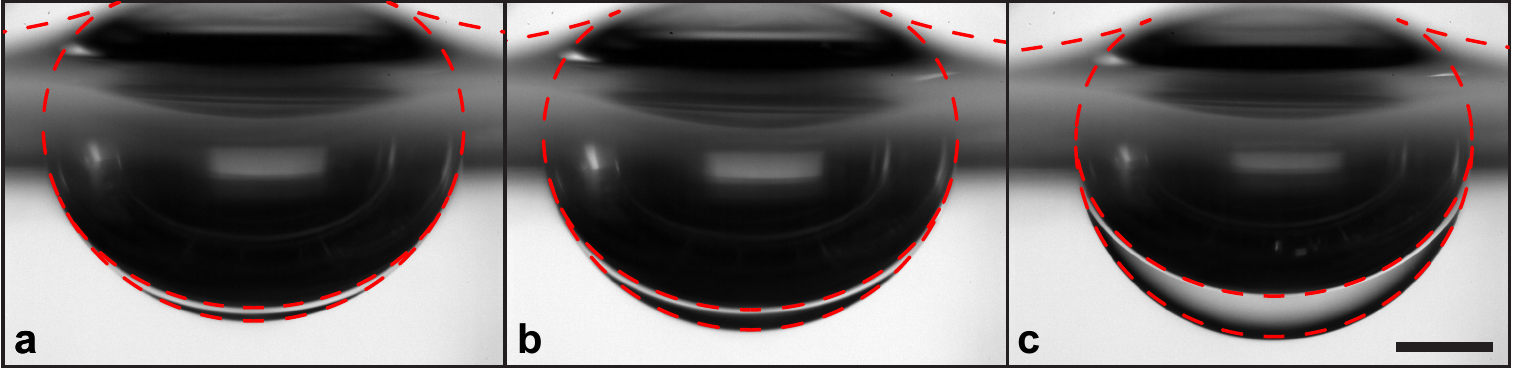}
    \caption{\textbf{Comparison of the numerically calculated static shape of oil-coated bubbles (red dashed curves) with experimental images.} The static shapes of the bubbles with $\mu_o$ = 1.8 mPa s and $\psi_o = 2.3\%$~(\textbf{a}), $4.3\%$~(\textbf{b}), and $12.6\%$~(\textbf{c}) resting at the free surface prior to bursting are well captured by the numerical solutions with the fluid properties listed in Extended Data Table \ref{tab:example}. The scale bar represents 1 mm.}
    \label{Num_Comp}
\end{extfigure*}

\begin{extfigure*}
    \centering
    \includegraphics[width=0.93\linewidth]{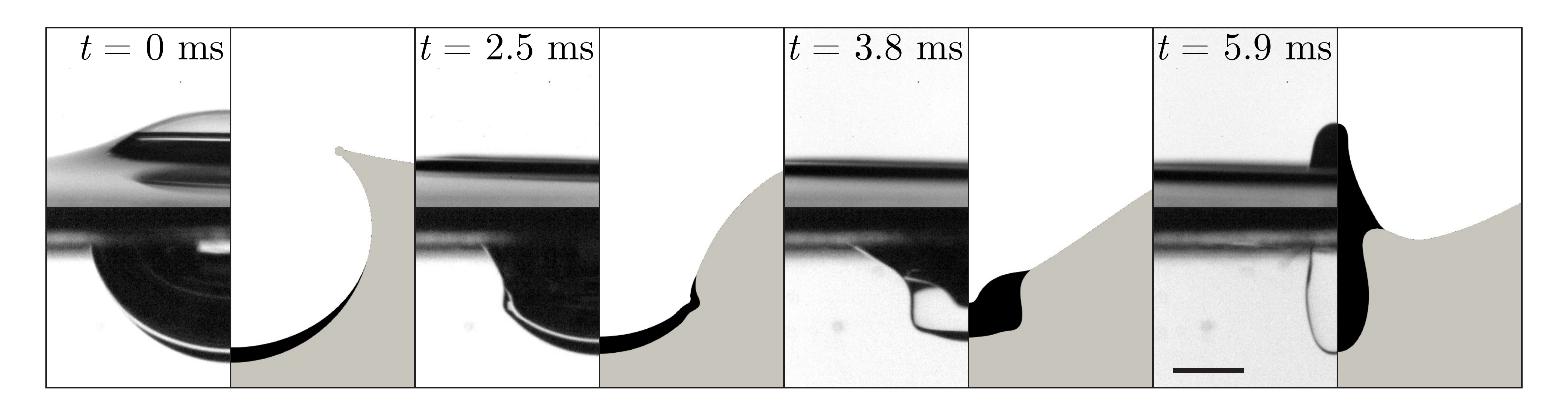}
    \caption{\textbf{Comparison of the experiment and simulation for oil-coated bubble bursting.} Left of each panel shows the experimental high-speed images of an oil-coated bubble bursting. Here $\mu_o =19$ mPa s, $\psi_o=4.2\%$. $t=0$ represents the instant when a hole nucleates in the bubble cap. Right of each panel shows the simulation snapshots of corresponding cavity shape. The white, black, and grey regimes denote air, oil, and water phases, respectively. The scale bar represents 1 mm. }
    \label{Comparison}
\end{extfigure*}

\begin{extfigure*}
    \centering
    \includegraphics[width=0.9\linewidth]{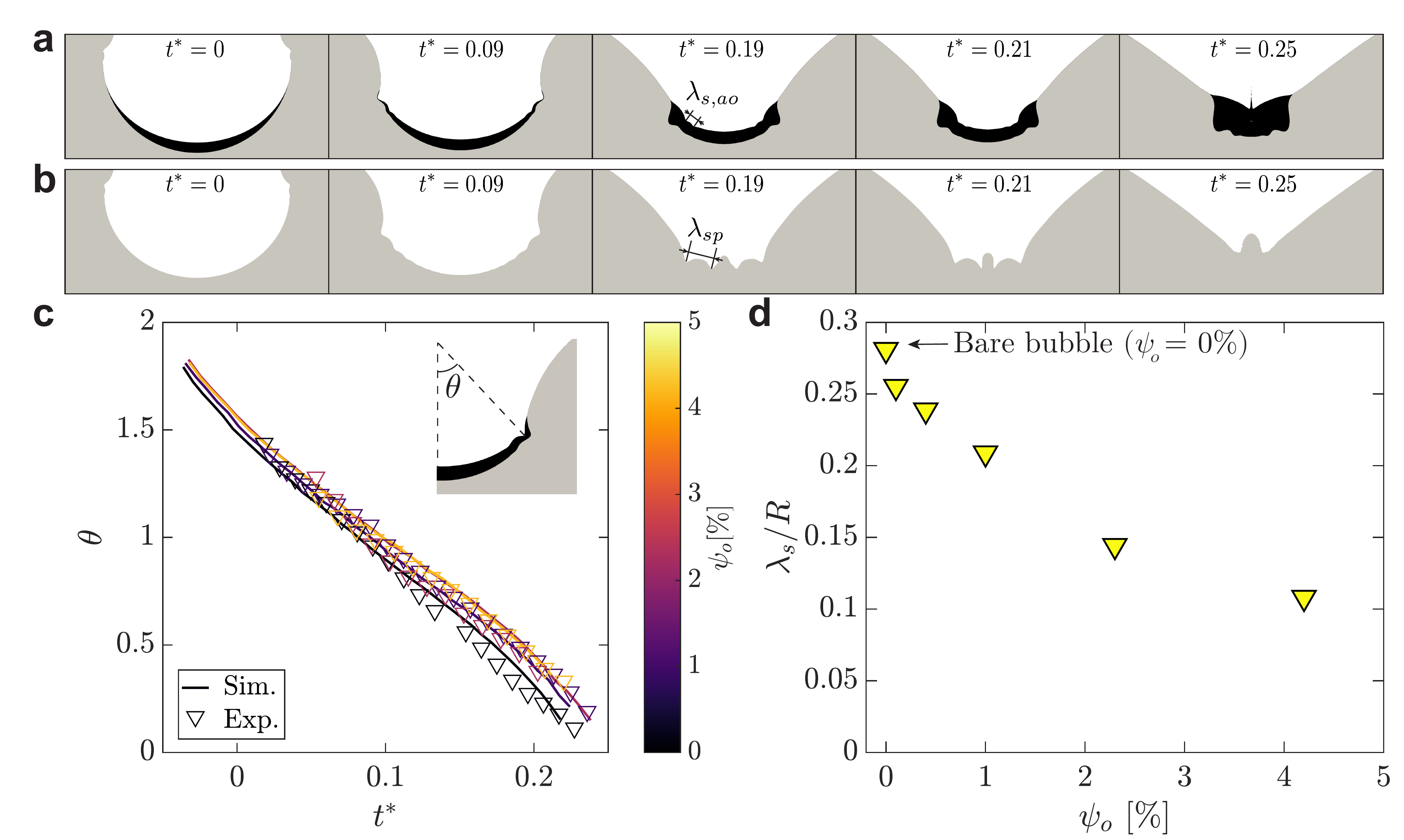}
    \caption{\textbf{Characterization of the SW propagation for oil-coated bubble bursting.} \textbf{a-b}, Capillary wave propagation during the bursting of an oil-coated bubble with $\mu_o=1.8$ mPa s and $\psi_o=4.2\%$ (\textbf{a}) and a bare bubble (\textbf{b}). White, black and grey colors represent air, oil, and water phases, respectively. The bubble radius $R = 2$ mm. The scale bar represents 1 mm. \textbf{c}, Angular wave position $\theta$ as a function of $t^*$ for oil-coated bubble bursting with $R =2$ mm and $\mu_o=4.6$ mPa s at different $\psi_o$. \textbf{d}, Dimensionless SW wavelength $\lambda_s/R$ as a function of $\psi_o$ at $\theta=\pi/6$ for oil-coated bubbles with $\mu_o$ = 4.6 mPa s. }
    \label{fig:WaveProp}
\end{extfigure*}

\begin{extfigure*}
    \centering
    \includegraphics[width=1\linewidth]{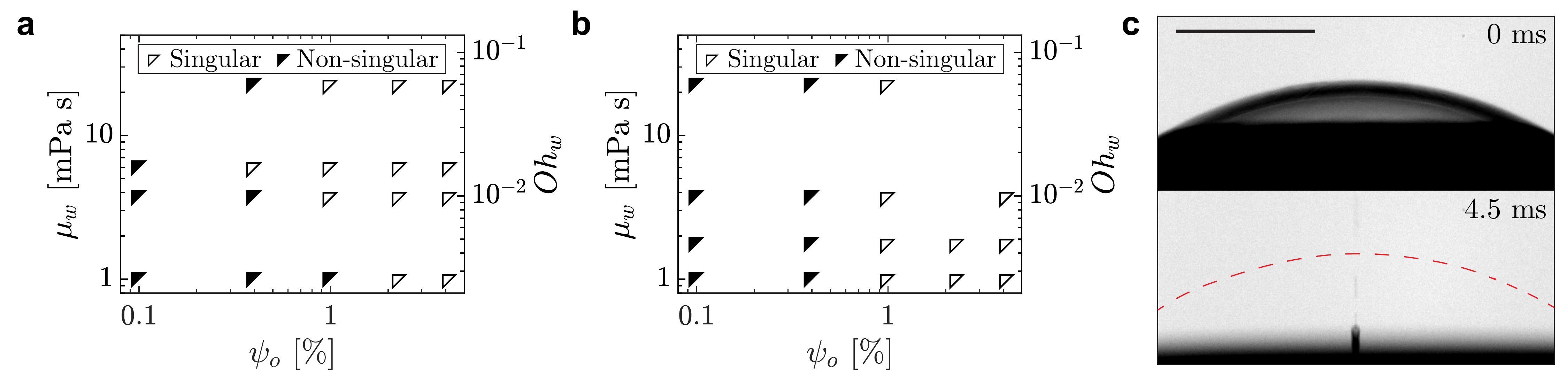}
    \caption{\textbf{Bubble bursting jet with different bulk liquid viscosities.} \textbf{a-b}, Regime map of jet singularity regarding oil fraction $\psi_o$ and bulk liquid viscosity $\mu_w$ (or $Oh_w=\mu_w/\sqrt{\rho_w R \gamma_{wa}}$), with an coating oil viscosity of 1.8 mPa s (\textbf{a}) and 4.6 mPa s (\textbf{b}). \textbf{c}, Experimental snapshots of a singular jet produced by bubble bursting with $\mu_w=$ 22.5 mPa s, $\mu_o=$ 4.6 mPa s, and $\psi_o=1.0\%$. The red dashed line marks the bubble cap before rupturing. The scale bar represents 1 mm. }
    \label{Bulk}
\end{extfigure*}
\end{document}